\documentclass[12pt]{article}

\pdfoutput=1

\usepackage{amssymb,amsmath}
\usepackage{amsfonts}
\usepackage[english ]{babel}
\usepackage[utf8]{inputenc} 
\usepackage{amssymb, subfigure,graphicx,graphics}
\usepackage{bbm}
\usepackage{authblk}
\usepackage{multirow}
\usepackage{arydshln}
\usepackage[usenames]{color}
\usepackage{pstricks,pst-node,pst-text,pst-3d, nopageno}
\usepackage{enumerate}
\usepackage{fullpage}

\newtheorem{teorema}{Theorem}
\newtheorem{lema}{Lemma}
\newtheorem{observacao}{Remark}
\newtheorem{proposicao}{Proposition}
\newtheorem{corolario}{Corollary}

\newtheorem{claim}{Claim}

\newenvironment{prova}[1][]{\par \noindent {\bf Proof:#1} 
}{ \hfill$\Box$} 

\newcommand{\R}{\mathbb{R}}
\newcommand{\Z}{\mathbb{Z}}

\newcommand{\N}{\mathbb{N}}

\newcommand{\asym}{\alpha ( \overline{W}^{n}_{p} )}
\newcommand{\afrac}{\left \lfloor \frac{n}{p}  \right\rfloor }
\newcommand{\antiweb}{\overline{W}^{n}_{p}}
\newcommand{\web}{W^{n}_{p}}
\newcommand{\asymweb}{\alpha(W^{n}_{p})}
\newcommand{\lc}{\left \lceil}
\newcommand{\rc}{\right \rceil}
\newcommand{\lf}{\left \lfloor}
\newcommand{\rf}{\right \rfloor} 

\begin{document} 

\title{Optimal $k$-fold colorings of webs and antiwebs\thanks{A short version of this paper was presented at {\em {S}imp\'osio {B}rasileiro de {P}esquisa {O}peracional, 2011.} This work is partially supported by a CNPq/FUNCAP Pronem project.}}

 \author[a]{Manoel Camp{\^e}lo\thanks{Partially supported by CNPq-Brazil. {\tt mcampelo@lia.ufc.br}}}
\author[b]{Ricardo C. Corr\^{e}a\thanks{\tt correa@lia.ufc.br}}
\author[c]{Phablo F. S. Moura\thanks{Partially supported by CNPq-Brazil. Most of this work
was done while the author was affiliated to Universidade Federal do Cear\'a. {\tt phablo@ime.usp.br}}}
\author[b]{Marcio C. Santos\thanks{Partially supported by Capes-Brazil. {\tt marciocs5@lia.ufc.br}}}

\affil[a]{\small Universidade Federal do Cear\'a,
Departamento de Estat\'\i stica e Matem\'atica Aplicada,
Campus do Pici, Bloco 910,
60440-554 Fortaleza - CE, Brazil}
\affil[b]{\small Universidade Federal do Cear\'a,
Departamento de Computa\c c\~ao,
Campus do Pici, Bloco 910,
60440-554 Fortaleza - CE, Brazil}
\affil[c]{\small Universidade de São Paulo,
Instituto de Matem\'atica e Estat\'{\i}stica,
Rua do Mat\~ao 1010, 05508-090 S\~ao Paulo - SP, Brazil}

\maketitle

\begin{abstract}
A $k$-fold $x$-coloring of a graph is an assignment of (at least) $k$ distinct colors
 from the set $\{1,2, \ldots, x \}$ to each vertex such that any two
adjacent vertices are assigned disjoint sets of colors. The smallest
number $x$ such that $G$ admits a $k$-fold $x$-coloring is the $k$-th
chromatic number of  $G$, denoted by $\chi_{k} (G)$. We determine the exact value of
this parameter when $G$ is a web or an antiweb. 
Our results generalize the known corresponding results for odd cycles and imply necessary and
sufficient conditions under which $\chi_k(G)$ attains its lower and upper bounds
based on the clique, the fractional chromatic and the chromatic numbers.
Additionally, we extend the concept of $\chi$-critical graphs to
$\chi_k$-critical graphs. We identify the webs and antiwebs having this
property, for every integer $k\geq 1$.

\end{abstract}

\section{Introduction}

For any integers $k \geq 1$ and $x\geq 1$, a \emph{$k$-fold $x$-coloring} of a graph is an assignment of (at least) $k$
distinct colors to each vertex from the set $\{1,2, \ldots, x \}$ such that any two adjacent vertices 
are assigned disjoint sets of colors \cite{GuaYue10,Sta76}. 
Each color used in the coloring defines what is called a {\em stable set} of the graph, 
{\it i.e.} a subset of pairwise nonadjacent vertices. We say that a graph $G$ is \emph{$k$-fold $x$-colorable} if $G$ admits a 
$k$-fold $x$-coloring. The smallest number $x$ such  that a graph $G$
is $k$-fold $x$-colorable is called the \emph{$k$-th chromatic number of  $G$} and
is denoted by $\chi_{k} (G)$ \cite{Sta76}. Obviously, $\chi_{1} (G) = \chi (G)$ is  the
conventional \emph{chromatic number of $G$}. 
This variant of the conventional graph coloring was
introduced in the context of radio frequency assignment problem \cite{Nar02,
Rob79}. Other applications include
scheduling problems, bandwidth allocation in radio networks, fleet maintenance and traffic phasing problems
\cite{BarGaf89,HalKor02,KlaMorPer08,OpsRob81}.

\begin{figure}[htb]
\centering 
\subfigure[$W^8_3$]{ \label{fig:web} \includegraphics[scale = 0.6]{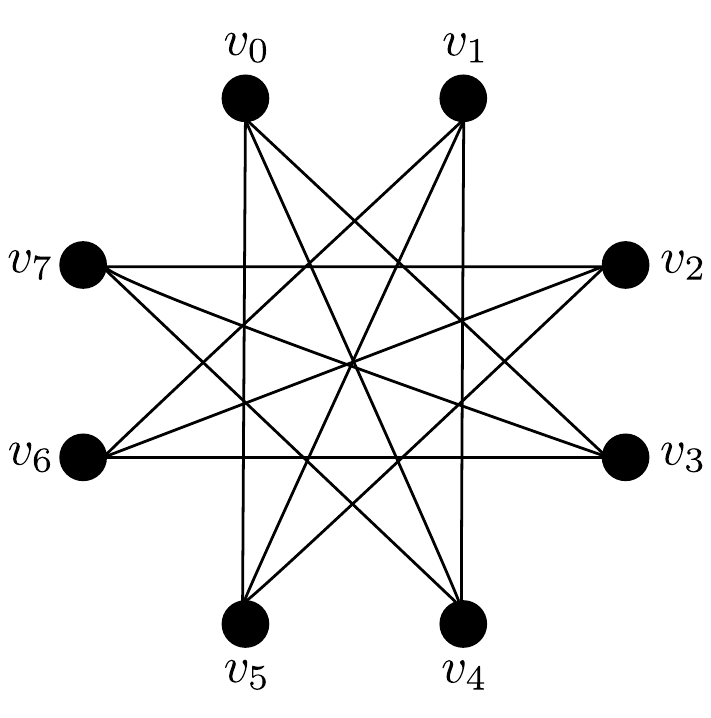}}
\qquad
\subfigure[$\overline{W}^8_3$]{\label{fig:antiweb} \includegraphics[scale = 0.6]{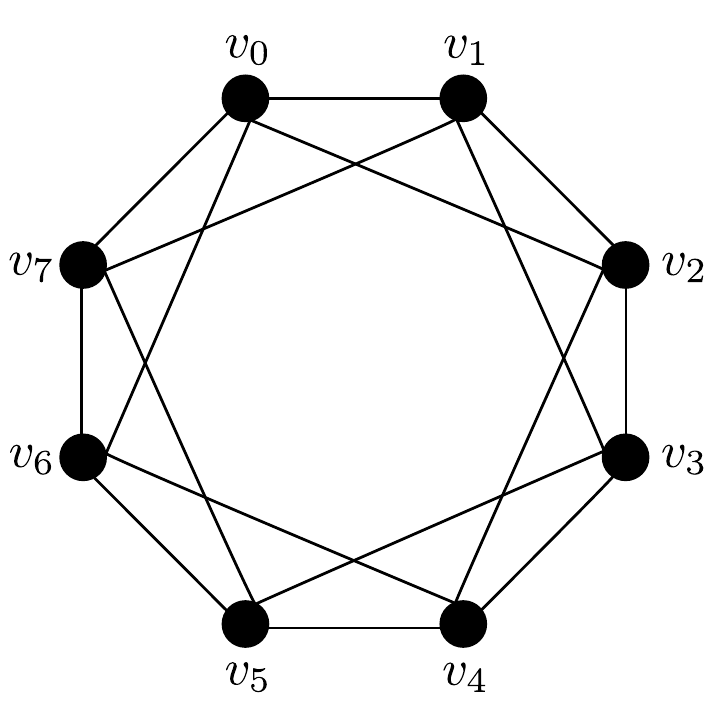}}
\caption{Example of a web and an antiweb. \label{fig:example}}
\end{figure}

Let $n$ and $p$ be integers such that $p \geq 1$ and $n \geq 2p$. As defined by Trotter, the {\em web 
$\web$} is the graph whose vertices can be labelled as $\{ v_{0},
v_{1}, \ldots, v_{n-1} \}$ in such a way that its edge set is $\{ v_{i}v_{j} \mid
p \leq |i -j| \leq n-p \}$~\cite{Tro75}. The {\em antiweb
$\antiweb$} is defined as the complement of $\web$. 
Examples are depicted in Figure~\ref{fig:example}, where the vertices are named according to an appropriate labelling (for the sake of convenience, we often name the vertices in this way in the remaining of the text).
We observe that these definitions are interchanged in some references (see \cite{PecWag06, Wag04}, for instance).
Webs and antiwebs form a class of graphs that play an important role in the context of stable sets and vertex coloring problems \cite{CheVri02A, CheVri02B, EOSV08, GalSas97, GilTro81, OriSta08, Pal10, PecWag06, Wag04}.

In this paper, we derive a closed formula for
the $k$-th chromatic number of webs and antiwebs. More specifically, we prove that
$\chi_{k}(\web ) = \lc \frac{kn}{p} \rc$  and $\chi_{k}(\antiweb ) = \lc \frac{kn}{\lf \frac{n}{p}
\rf} \rc$, for every $k\in \N$, thus generalizing similar results for odd cycles \cite{Sta76}. 
The denominator of each of these formulas is the size of the largest stable set in the corresponding graph, {\it i.e.} the {\em stability number} of the graph \cite{Tro75}. Besides this direct relation with the stability number, we also relate the $k$-th chromatic number of webs and antiwebs with other parameters of the graph, such as the clique, chromatic and fractional chromatic numbers. Particularly, we derive necessary and sufficient conditions under which the classical bounds given by these parameters are tight.

In addition to the value of $k$-th chromatic number, we also provide optimal $k$-fold colorings of $\web$ and $\antiweb$.
Based on the optimal colorings, we analyse when webs and antiwebs are critical with respect to this parameter. 
A graph $G$ is said to be {\em $\chi$-critical} if $\chi(G - v) < \chi(G)$, for all $v \in V(G)$. 
An immediate consequence of this definition is that if $v$ is a vertex of a $\chi$-critical
graph $G$, then there exists an optimal $1$-fold coloring of $G$ such that the color of $v$ is not 
assigned to any other vertex. Not surprisingly, $\chi$-critical subgraphs of $G$ play an important 
role in several algorithmic approaches to vertex coloring. For instance, they are the core 
of the reduction procedures of the heuristic of \cite{HerHer02} as well as they give facet-inducing inequalities 
of vertex coloring polytopes explored in cutting-plane methods \cite{CamCorFro04,HanLabSch09,MenDia08}. From this algorithmic 
point of view, odd holes and odd anti-holes are (along with cliques) the most widely used 
$\chi$-critical subgraphs. It is already been noted that not only odd holes or odd anti-holes, but 
also $\chi$-critical webs and antiwebs give facet-defining inequalities \cite{CamCorFro04,Pal10}. 

We extend the concept of $\chi$-critical graphs to $\chi_k$-critical graphs in a straightforward way. Then, 
we characterize $\chi_k$-critical webs and antiwebs, for any integer $k\geq 1$. The characterization crucially depends on the greatest common divisors between $n$ and $p$ and between $n$ and the stability number (which are equal for webs but may be different for antiwebs).  
Using the B\'ezout's identity, we show that there exists $k\geq 1$ such that $\web$ is $\chi_k$-critical if, and only if, $\gcd(n,p) = 1$. Moreover, when this condition holds, we determine all values of $k$ for which $\web$ is $\chi_k$-critical. Similar results are derived for $\antiweb$, where the condition $\gcd(n,p) = 1$ is replaced by $\gcd(n,p) \neq p$. As a consequence, we obtain that a web or an antiweb is $\chi$-critical if, and only if, the stability number divides $n-1$. Such a characterization is trivial for webs but it was still not known for antiwebs \cite{Pal10}.
More surprising, we show that being $\chi$-critical is also a sufficient for a web or an antiweb to be $\chi_k$-critical for all $k\geq 1$.

Throughout this paper, we mostly use notation and definitions consistent with what is
generally accepted in graph theory. Even though, let us
set the grounds for all the notation used from here on. Given a graph $G$, $V(G)$ and $E(G)$
stand for its set of vertices and edges, respectively. The simplified notation
$V$ and $E$ is prefered when the graph $G$ is clear by the context.
The complement of $G$ is written as $\overline{G} = (V, \overline{E})$.  
The edge defined by vertices $u$ and $v$ is denoted by $uv$. 

As already mentioned, a set $S \subseteq V(G)$ is said to be a {\em stable set} if all vertices 
in it are pairwise non-adjacent in $G$, i.e. $uv \not \in E$ $\forall u,v \in
S$. The {\em stability number} $\alpha (G)$ of $G$ is the size of the largest 
stable set of $G$. Conversely, a {\em clique} of $G$ is a subset $K\subseteq V(G)$ of
pairwise adjacent vertices. The {\em clique number} of $G$ is the size of the largest clique and is denoted 
by $\omega(G)$. For the ease of expression, we frequently refer to the graph itself as being a clique (resp. stable set) if its vertex set is a clique (resp. stable set). 
The \emph{fractional chromatic number of $G$}, to be denoted $\bar \chi(G)$,  is
the infimum of $\frac{x}{k}$ among the $k$-fold $x$-colorings \cite{SchUll97}.
It is known that $\omega(G) \leq \bar \chi(G) \leq \chi(G)$ and $\frac{n}{\alpha(G)} \leq \bar \chi(G)$~\cite{SchUll97}. A graph $G$ is {\em
perfect} if $\omega(H)=\chi(H)$, for all induced subgraph $H$ of $G$.

A {\em chordless cycle} of length $n$ is a graph $G$ such that $V=\{v_1,v_2,\ldots,v_{n}\}$ and $E=\{v_iv_{i+1}:i=1,2,\ldots,n-1\}
\cup \{v_1v_n\}$. A {\em hole} is a chordless cycle of length at least four.
An {\em antihole} is the complement of a hole. Holes and antiholes are odd or even according to the
parity of their number of vertices. Odd holes and odd antiholes are minimally
imperfect graphs \cite{ChuRobSeyTho06}.
Observe that the odd holes and odd anti-holes are exactly the webs $W_{\ell}^{2\ell+1}$ and $W_{2}^{2\ell+1}$, for some 
integer $\ell \geq 2$, whereas the cliques are exactly the webs $W^n_1$.

In the next section, we present general lower and upper bounds for the $k$-th
chromatic number of an arbitrary simple graph. The exact value of this parameter
is calculated for webs (Subsection~\ref{subsec:web}) and antiwebs
(Subsection~\ref{subsec:antiweb}). Some consequences of this result are 
presented in the following sections.  In Section~\ref{sec:tight}, we relate the $k$-th
chromatic number of webs and antiwebs to their clique, integer and fractional chromatic numbers.
In particular,
we identify which webs and antiwebs achieve the bounds given in Section~\ref{sec:bounds} and those for which these bounds are strict.
The definitions of $\chi_k$-critical and $\chi_*$-critical graphs are introduced
in  Section~\ref{sec:critical}, as a natural extension of the concept of
$\chi$-critical graphs.  Then, we identify all webs and antiwebs 
that have these two properties.

\section{Bounds for the $k$-th chromatic number of a graph}\label{sec:bounds}

Two simple observations lead to lower and upper bounds for the $k$-th
chromatic number of a graph $G$. On one hand, every vertex of a clique of $G$ must receive $k$ colors different from any color assigned to the other vertices of the clique. On the other hand, a $k$-fold 
coloring can be obtained by just replicating an $1$-fold coloring $k$ times. Therefore, we get the following bounds which are tight, for 
instance, for perfect graphs.

\begin{lema} \label{lem:up}
For every $k\in \N$,  $\omega(G)\leq \bar \chi(G)\leq \frac{\chi_{k}(G)}{k}\leq \chi(G)$.
\end{lema}

Another lower bound is related to the stability number, as follows. 
The lexicographic product of a graph $G$ by a graph $H$ is the graph that we obtain
by replacing each vertex of $G$ by a copy of $H$ and adding all edges
between two copies of $H$ if and only if the two replaced vertices of $G$ were
adjacent. More formally, the {\em lexicographic product} $G \circ H$ is a graph such that:
\begin{enumerate}
  \item the vertex set of $G \circ H$ is the cartesian product $V(G) \times
  V(H)$; and
  \item any two vertices $(u,\hat{u})$ and $(v,\hat{v})$ are adjacent in $G
  \circ H$ if and only if either $u$ is adjacent to $v$, or $u = v$ and
  $\hat{u}$ is adjacent to $\hat{v}$
\end{enumerate}
As noted by Stahl, another way to interpret the $k$-th chromatic
number of a graph $G$ is in terms of $\chi(G \circ K_{k})$, where $K_{k}$ is a clique with
$k$ vertices \cite{Sta76}. It is easy to see that a $k$-fold $x$-coloring of $G$
is equivalent to a $1$-fold coloring of $G \circ K_{k}$ with $x$ colors. Therefore,
$\chi_{k} (G) = \chi (G \circ K_{k})$. Using this equation we can trivially derive the following lower bound
for the $k$-th chromatic number of any graph. 

\begin{lema}\label{lema:lex}
For every graph $G$ and every $k \in \N$, $\chi_{k} (G) \geq \left\lceil 
\frac{kn}{\alpha(G)} \right\rceil$.
\end{lema}
\begin{prova}
If $H_{1}$ and $H_{2}$ are two graphs, then $\alpha(H_{1} 
\circ H_{2}) = \alpha(H_{1}) \alpha(H_{2})$ \cite{GelSta75}.
Therefore, $\alpha(G \circ K_{k}) = \alpha(G) \alpha(K_{k}) = \alpha(G)$.
We get $\chi_{k} (G) = \chi(G \circ K_{k}) \geq \left\lceil
\frac{kn}{\alpha (G\circ K_{k})} \right\rceil = \left\lceil \frac{kn}{\alpha
(G)} \right\rceil$.
\end{prova}

Next we will show that the lower bound given by Lemma~\ref{lema:lex} is tight for two classes of graphs,
namely webs and antiwebs. Moreover, some graphs in these classes also achieve
the lower and upper bounds stated by Lemma~\ref{lem:up}.
 
\section{The $k$-th chromatic number of webs e antiwebs}\label{sec:chik}

In the remaining, let $n$ and $p$ be integers such that $p \geq 1$ and $n \geq
2p$ and let $\oplus$ stand for addition modulus $n$, i.e. $i \oplus j = (i+j) 
\mod n$ for $i,j \in \Z$. Let $\N$ stand for the set of natural numbers ($0$ excluded). The following known results will be used later.

\begin{lema}[Trotter~\cite{Tro75}] \label{lema:ao}
$\asym =\omega(\web)= \afrac$ and $\alpha(\web)=\omega(\antiweb)=p$.
\end{lema}

\begin{lema}[Trotter~\cite{Tro75}] \label{lema:trotter}
Let $n'$ and $p'$ be integers such that $p' \geq 1$ and $n' \geq
2p'$. The web $W^{n'}_{p'}$ is a subgraph of $\web$ if, and only if, $np'\geq n'p$ and $n(p'-1)\leq n'(p-1)$. 
\end{lema}

\subsection{Web}\label{subsec:web}

We start by defining some stable sets of $\web$.
For each integer $i\geq 0$, define the following sequence of integers:
\begin{equation} \label{eq:Sweb}
 S_i = \langle i\oplus 0,i\oplus 1,\ldots, i\oplus (p-1) \rangle
\end{equation}
\begin{lema} \label{lem:web}
For every integer $i\geq 0$, $S_i$ indexes a maximum stable set of $\web$.
\end{lema}
\begin{prova}
By the symmetry of $\web$, it suffices to consider the sequence $S_0$. Let
$j_1$ and $j_2$ be in $S_0$. Notice that $|j_1-j_2| \leq p-1 < p$. Then, $v_{j_1}v_{j_2} \notin E(\web)$, which proves that $S_0$ indexes an independent set with cardinality $p=\alpha(\web)$.
\end{prova}

Using the above lemma and the sets $S_{i}$, we can now calculate the $k$-th
chromatic number of $\web$. The main ideia is to build a cover of the graph by
stable sets in which each vertex of $\web$ is covered at least $k$ times.

\begin{teorema} \label{teo:web}
For every $k\in \N$, $\chi_{k}(\web ) = \lc \frac{kn}{p} \rc = \lc \frac{kn}{\asymweb} \rc$.
\end{teorema}
\begin{prova}
By Lemma~\ref{lema:lex}, we only have to show that $\chi_{k}(\web ) \leq \lc
\frac{kn}{p}\rc$, for  an arbitrary $k\in \N$. For this purpose, we show that
$\Xi(k)  = \langle S_{0}, S_{p}, \ldots,S_{(x - 1)p} \rangle$ gives a $k$-fold 
$x$-coloring of $\web$, with $x=\lc \frac{kn}{p}\rc$. We have that 
\begin{multline*}
\Xi(k) = \left\langle \underbrace{ 0 \oplus 0 , 0 \oplus 1,\ldots, 0 \oplus
p-1}_{S_0},\underbrace{p  \oplus 0 ,\ldots, p \oplus (p-1)}_{S_p},\ldots, \right. \\
\left. \underbrace{ (x-1)p \oplus 0,\ldots, (x-1)p \oplus (p-1)}_{S_{(x-1)p}}
\right\rangle.
\end{multline*}
Since the first element of $S_{(\ell+1)p}$, $0 \leq \ell < x-1$, is the
last element of $S_{\ell p}$ plus $1$ (modulus $n$), we have that $\Xi(k)$ is a sequence 
(modulus $n$) of integer numbers starting at $0$. Also, it has $\lc \frac{kn}{p}
\rc p\geq kn$ elements. Therefore, each element between $0$ and $n-1$ appears at
least $k$ times in $\Xi(k)$. By Lemma~\ref{lem:web}, this means that $\Xi(k)$
gives  a $k$-fold $\lc \frac{kn}{p}\rc$-coloring of $\web$, as
desired.
\end{prova}

\subsection{Antiweb}\label{subsec:antiweb}

\begin{figure}
\centering
\subfigure[$\overline{W}^{10}_3$. \label{subfig:aw1}]{\includegraphics[scale=0.6]{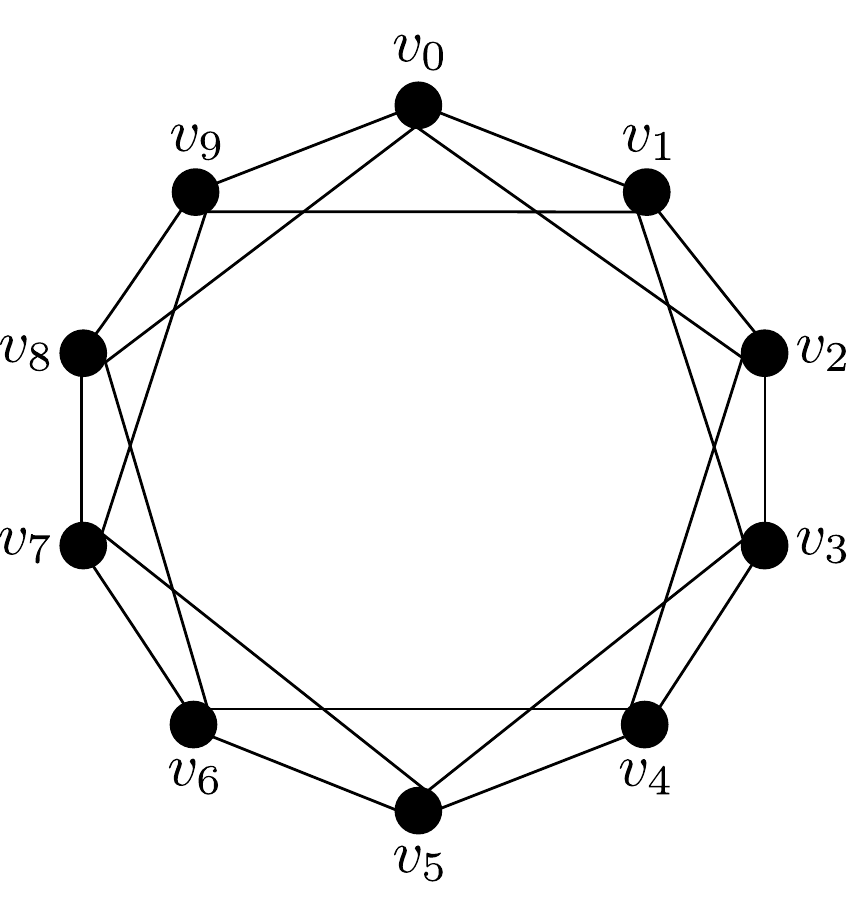}}
\qquad
\subfigure[$C(1)$ in blue, $C(2)$ in red. \label{subfig:aw2}]{
$
\begin{array}[b]{lcccc:ccc:c}
          & S_0 & S_1 & S_2 & S_3 & S_4 & S_5 & S_ 6& S_7\\ \hline
A(\ell,0) & \blue 0  & \blue 1 & \blue 2 & \blue 3 & \red  4 & \red 5 & \red  6 & \cdots \\ 
A(\ell,1) & \blue 4  & \blue 5 & \blue 6 & \red  7 & \red  8 & \red 9 &   0 & \cdots \\ 
A(\ell,2) & \blue 7  & \blue 8 & \blue 9 & \red  0 & \red  1 & \red 2 &  \red 3 & \cdots \\ \hline
\\
          & \multicolumn{4}{c:}{\ell =1}   &    &     & \\ \cdashline{2-5}      
          & \multicolumn{4}{c}{}   \\
          & \multicolumn{4}{r}{\ell =2 \:\:\:}    &    &     & \\ \cdashline{2-8}
\end{array}
$
}
 \caption{Example of a 2-fold 7-coloring of $\overline{W}^{10}_ 3$. Recall that $\alpha(\overline{W}^{10}_3)=3$.}
\label{fig:antiweb1}
 \end{figure}

As before, we proceed by determining stable sets of $\antiweb$
that cover each vertex at least $k$ times. Now, we need to be more judicious in the choice of
the stable sets of $\antiweb$. We start by defining the following sequences (illustrated in Figure~\ref{fig:antiweb1}):
\begin{eqnarray}
S_0 &=& \left\langle \lc t{\textstyle \frac{n}{\asym}} \rc: \; t=0,1,\ldots,
\asym-1 \right\rangle \\ \nonumber
 S_i &=& \langle j \oplus 1: \; j\in S_{i-1} \rangle, \quad
i \in \N\\ \label{eq:Santiweb}
 &=& \langle j \oplus i: \; j\in S_{0} \rangle, \quad i \in \N. \nonumber
\end{eqnarray}

We claim that each $S_i$ indexes a maximum stable set of $\antiweb$. This will be shown with the help of the following lemmas.

\begin{lema}\label{lemaTetoPiso}
If $x,y \in \R$ and $x \geq y$, then  $\lf x - y \rf \leq \lc x \rc - \lc y \rc \leq \lc x-y \rc$.
\end{lema}
\begin{prova}
It is clear that $x-\lc x \rc \leq 0$ and $\lc y \rc -y < 1$. By summing up these inequalities, we get $\lf x-y +\lc y \rc - \lc x \rc \rf \leq 0$.  Therefore,  $\lf x - y \rf \leq \lc x \rc - \lc y \rc$. To get the second inequality, recall that $\lc x-y \rc + \lc y \rc \geq \lc x-y+y \rc =\lc x \rc$.
\end{prova}

\begin{lema}\label{lemaPK}
For every antiweb $\antiweb$ and  every integer $k \geq 0$, $\lf \frac{nk}{\asym}\rf \geq
pk$.
\end{lema}
\begin{prova}
 Since $\asym = \afrac$, we have that $\frac{n}{p}
\geq \asym$, which implies $\frac{nk}{\asym} \geq pk$. Since $pk$ is integer, the result follows.
\end{prova}

\begin{lema}\label{lemaP}
For  $\antiweb$ and every integer $\ell \geq 1$, $\lc \frac{\ell n}{\asym}\rc -
\lc \frac{(\ell-1)n}{\asym}\rc \geq p$. 
\end{lema}
\begin{prova}
By Lemma \ref{lemaTetoPiso}, we get 
$$
\lc \frac{\ell n}{\asym} \rc -
\lc \frac{(\ell-1)n}{\asym}\rc \geq \lf \frac{\ell n}{\asym} -
\frac{(\ell-1)n}{\asym} \rf = \lf \frac{n}{\asym} \rf.$$ 
The statement then follows from Lemma \ref{lemaPK}.
\end{prova}

We now get the counterpart of Lemma~\ref{lem:web} for antiwebs.
\begin{lema} \label{lema:Santiweb}
For every integer $i\geq 0$, $S_{i}$ indexes a maximum stable set of
$\antiweb$.
\end{lema}
\begin{prova}
By the symmetry of an antiweb and the definition of the $S_{i}$'s, it suffices to show the claimed result for $S_{0}$. Let 
$j_1$ and $j_2$ belong to $S_{0}$. We have to show that $p\leq |j_1-j_2| \leq n-p$. For the upper bound, note that 
$$|j_1-j_2| \leq \lc 
  \frac{(\asym -1)n}{\asym} \rc =  \lc n-\frac{n}{\asym} \rc.$$
Lemma \ref{lemaPK} implies that this last term is no more than 
$\lc n-p \rc$, that is, $n-p$. On the other hand, 
$$
|j_1-j_2| \geq \min \limits_{\ell
\geq 1} \left(\lc \frac{\ell n}{\asym} \rc - \lc \frac{(\ell-1)n}{\asym} \rc \right). 
$$
By Lemma \ref{lemaP}, it follows that $|j_1 - j_2| \geq p$. Therefore, $S_0$ indexes an independent set of cardinality $\asym$.
\end{prova}

The above lemma is the basis to give the expression of $\chi_{k}(\antiweb)$. We proceed by choosing an appropriate
family of $S_i$'s and, then, we show that it covers each vertex at least $k$ times. We first consider
the case where $k \leq \asym$.

\begin{lema}\label{lemaUB}
Let be given positive integers $n$, $p$, and $k\leq \asym$. The index of each vertex of $\antiweb$ belongs to at least 
$k$ of the sequences $S_{0}, S_1,\ldots, S_{x(k)-1}$, where $x(k) = \lc \frac{kn}{\asym}\rc$.
\end{lema}
\begin{prova}
Let $\ell\in \{1,2,\dots,k\}$ and $t\in \{0,1,\ldots,{\asym-1}\}$. 
Define $A(\ell,t)$ as the sequence comprising the $(t+1)$-th elements of $S_0,S_1,\ldots,S_{x(\ell)-1}$, that is,
$$
A(\ell,t)=\left \langle \lc t{\textstyle \frac{n}{\asym}} \rc \oplus i: \;
i=0,1,\ldots, \lc {\textstyle \frac{\ell n}{\asym}}\rc  -1 \right \rangle.
$$
Since $\ell \leq \asym$, $A(\ell,t)$ has $\lc \frac{\ell n}{\asym} \rc$ distinct elements. Figure~\ref{fig:antiweb1} illustrates these sets for $\overline{W}^{10}_3$.

Let $B(\ell,t)$ be the subsequence of $A(\ell,t)$ formed by its first 
$\lc \frac{(\ell+t)n}{\asym}\rc - \lc \frac{tn}{\asym} \rc \leq \lc \frac{\ell n}{\asym} \rc$ elements (the inequality comes 
from Lemma \ref{lemaTetoPiso}). In Figure~\ref{subfig:aw2}, $B(1,t)$ relates to the numbers in blue whereas $B(2,t)$ comprises the numbers in blue and red. 
Notice that $B(\ell,t)$ comprises consecutive integers (modulus $n$), starting at $\lc {\textstyle \frac{tn}
{\asym}} \rc \oplus 0$ and ending at $\lc {\textstyle \frac{(\ell+t)n}{\asym}} \rc \oplus (-1)$. 
Consequently, $B(\ell, t) \subseteq B(\ell + 1, t)$.

Let $C(1,t)=B(1,t)$ and $C(\ell+1,t)=B(\ell+1,t)\setminus B(\ell,t)$, for $\ell<k$.  Similarly to $B(\ell,t)$, $C(\ell,t)$ comprises 
consecutive integers (modulus $n$), starting at $\lc {\textstyle \frac{(\ell+t-1)n}{\asym}} \rc \oplus 0$ and ending at $\lc 
{\textstyle \frac{(\ell+t)n}{\asym}} \rc \oplus (-1)$. Observe that the first element of $C(\ell,t+1)$ is the last element of $C(\ell,t)$ 
plus $1$ (modulus $n$). Then, $C(\ell)=\langle C(\ell,0),C(\ell,1),\ldots,C(\ell,\asym-1) \rangle$ is a sequence of consecutive integers (modulus $n$) 
starting at the first element of $C(\ell,0)$, that is $\lc {\textstyle \frac{(\ell-1) n}{\asym}} \rc \oplus 0$, and ending at the last 
element of $C(\ell,\asym-1)$, that is 
$$
\lc {\textstyle \frac{(\asym+\ell-1)n}{\asym}} \rc \oplus (-1)=\lc {\textstyle \frac{(\ell-1) 
n}{\asym}} \rc \oplus (-1).$$
This means that $C(\ell)\equiv \langle 0,1,\ldots,n-1\rangle$. Therefore, for each $\ell=1,2,\ldots,k$, $C(\ell)$ covers 
every vertex once. Consequently, every vertex is covered $k$ times by $C(1),C(2),\ldots,C(k)$, and so is covered at least $k$ times by    
$S_0,S_1,\ldots,S_{x(k)-1}$. 
\end{prova}

Now we are ready to prove our main result for antiwebs.

\begin{teorema} \label{teo:antiweb}
For every $k\in \N$, $\chi_{k}(\overline{W}^{n}_{p} ) = \lc
\frac{kn}{\alpha(\overline{W}^{n}_{p})} \rc$.
\end{teorema}
\begin{prova}
 By Lemma \ref{lema:lex}, we only need to show the inequality $\chi_{k}(\antiweb) \leq \lc  \frac{kn}{\asym} \rc$. 
 Let us write $k = \ell\asym + i$, for integers $\ell\geq 0$ and $0 \leq  i < \asym$. By lemmas~\ref{lema:Santiweb} and \ref{lemaUB}, it is straightforward that the stable sets $S_{0}, S_{1}, \ldots,
 S_{x-1}$, where $x = \lc \frac{in}{\asym} \rc $, induce an $i$-fold
 $x$-coloring of $\antiweb$. The same lemmas also give an $\asym$-fold $n$-coloring via sets 
 $S_{0}, \ldots, S_{n-1}$. One copy of the first coloring together with $\ell$ copies of the second one yield a $k$-fold coloring with $ \ell n + \lc  \frac{in}{\asym} \rc = \lc \frac{kn}{\asym} \rc$ colors. 
\end{prova}

\section{Relation with other parameters}\label{sec:tight}

The strict relationship between $\chi_k(G)$ and $\alpha(G)$ established for webs (Theorem~\ref{teo:web}) and anti-webs (Theorem~\ref{teo:antiweb}) naturally motivates a similar question with respect to other parameters of $G$ known to be related to the chromatic number. 
Particularly, we determine in this section when the bounds presented in Lemma~\ref{lem:up} are tight or strict.

\begin{proposicao}\label{prop:up}
Let $G$ be $\web$ or $\antiweb$ and $k\in \N$. Then, $\chi_{k}(G)= k\chi(G)$ if, and only if, 
$\gcd(n,\alpha(G))=\alpha(G)$ or $k< \frac{\alpha(G)}{\alpha(G)-r}$, where $r=n \mod \alpha(G)$.
\end{proposicao}
\begin{prova}
By theorems~\ref{teo:web} and \ref{teo:antiweb}, $\chi_{k}(G)= k\chi(G)$ if, and only if, $\lc \frac{kn}{\alpha(G)} \rc = k \lc 
\frac{n}{\alpha(G)}\rc$, which is also equivalent to $\lc \frac{kr}{\alpha(G)} \rc = k \lc \frac{r}{\alpha(G)}\rc$. This equality trivially holds if $r=0$, that is, $\gcd(n,\alpha(G))=\alpha(G)$. In the complementary case, $\lc \frac{r}{\alpha(G)}\rc=1$ and, consequently, the equality is equivalent to $\frac{kr}{\alpha(G)}>k-1$ or still $k< \frac{\alpha(G)}{\alpha(G)-r}$. 
\end{prova}

\begin{proposicao}\label{prop:lo}
Let $G$ be $\web$ or $\antiweb$ and $k\in \N$. Then, $\chi_{k}(G)= k\omega(G)$ if, and only if, $\gcd(n,p)=p$.
\end{proposicao}
\begin{prova}
Let $s=n \mod p$. Using Lemma~\ref{lema:ao}, note that $n=\lf n/p\rf p + s =\omega(G)\alpha(G)+s$. By theorems~\ref{teo:web} and \ref{teo:antiweb}, we get 
$$\chi_{k}(G)= \lc \frac{kn}{\alpha(G)}\rc = k\omega(G)+ \lc \frac{ks}{\alpha(G)}\rc.$$ 
The result then follows from the fact that $s = 0$ if, and only if, $\gcd(n, p) = p$.
\end{prova}

As we can infer from Lemma~\ref{lema:ao}, if $p$ divides $n$, then so does $\alpha(\web)$ and $\alpha(\antiweb)$. Under such a condition, which holds for all perfect and some non-perfect webs and antiwebs, the lower and upper bounds given in Lemma~\ref{lem:up} are equal.
\begin{corolario} \label{cor:eq}
Let $G$ be $\web$ or $\antiweb$ and $k\in \N$. Then, $k\omega(G)=\chi_{k}(G)= k\chi(G)$ if, and only if, $\gcd(n,p)=p$.
\end{corolario}

On the other hand, the same bounds are always strict for some webs and antiwebs, including the minimally imperfect graphs. 
\begin{corolario}
Let $G$ be $\web$ or $\antiweb$. 
If $\gcd(n-1,\alpha(G))=\alpha(G)$ and $\alpha(G)>1$, then $\chi_{k}(G)< k\chi(G)$, for all $k>1$. Moreover, if $\gcd(n-1,p)=p$ and $p>1$, then $k\omega(G)< \chi_{k}(G)< k\chi(G)$, for all $k>1$.
\end{corolario}
\begin{prova}
Assume that $\gcd(n-1,\alpha(G))=\alpha(G)$ and $\alpha(G)\geq 2$. Then, $r:=n\mod \alpha(G)=1$ and $\frac{\alpha(G)}{\alpha(G)-r}\leq 2$. By Proposition~\ref{prop:up}, $\chi_{k}(G)< k\chi(G)$ for all $k>1$. To show the other inequality, assume that $\gcd(n-1,p)=p$ and $p>1$. Then, $\gcd(n,p)\neq p$. Moreover, $\alpha(\web)=p>1$ and $\alpha(\antiweb)=\frac{n-1}{p}>1$ so that $\gcd(n-1,\alpha(G))=\alpha(G)>1$. By the first part of this corollary and Proposition~\ref{prop:lo}, the result follows. 
\end{prova}

To conclude this section, we relate the fractional chromatic number and the $k$-th chromatic number. 
By definition, for any graph $G$, these parameters are connected as follows:
$$
\bar \chi(G)=\inf \left\{ \frac{\chi_k(G)}{k} \mid \, k\in \N \right\}.
$$
By theorems~\ref{teo:web} and \ref{teo:antiweb}, $\frac{\chi_k(G)}{k}\geq \frac{n}{\alpha(G)}$, for every $k\in \N$, and this bound is attained with $k=\alpha(G)$. This leads to
\begin{proposicao}\label{prop:chibar}
If $G$ is $\web$ or $\antiweb$, then $\bar \chi(G)= \frac{n}{\alpha(G)}$.
\end{proposicao}

Actually, the above expression holds for a larger class of graphs, namely
vertex transitive graphs \cite{SchUll97}. The following property readily follows in the
case of webs and antiwebs.

\begin{proposicao}
Let $G$ be $\web$ or $\antiweb$ and $k\in \N$. Then, $\chi_{k}(G)= k\bar \chi(G)$ if, and only if, 
$\frac{k\gcd(n,\alpha(G))}{\alpha(G)}\in \Z$.
\end{proposicao}
\begin{prova}
Let $\alpha=\alpha(G)$ and $g=\gcd(n,\alpha)$. By theorems~\ref{teo:web} and
\ref{teo:antiweb} and Proposition~\ref{prop:chibar}, $\chi_{k}(G)= k\bar \chi(G)$ if, and only if, $\frac{kn}{\alpha}\in \Z$. Since $n/g$ and $\alpha/g$ are coprimes, $\frac{kn}{\alpha}=\frac{k(n/g)}{\alpha/g}$ is integer if, and only if, $\frac{k}{\alpha/g}\in \Z$.
\end{prova} 

By the above proposition, given any web or antiweb $G$ such that $\alpha(G)$ does not divide $n$, there are always values of $k$ such that $\chi_{k}(G)= k\bar \chi(G)$ and values of $k$ such that $\chi_{k}(G)> k\bar \chi(G)$.

\section{$\chi_k$-critical web and antiwebs} \label{sec:critical}
We define a {\em $\chi_k$-critical} graph as a graph $G$ such that $\chi_k(G-v)<\chi_k(G)$, for all $v\in V(G)$. 
If this relation holds for every $k\in \N$, then $G$ is said to be {\em $\chi_*$-critical}. Now we investigate these properties for webs and antiwebs.
The analysis is trivial in the case where $p=1$ because $W^n_1$ is a clique. For the case where $p>1$, the following property will be useful.

\begin{lema} \label{lem:eq}
If $G$ is $\web$ or $\antiweb$ and $p>1$, then $\alpha(G-v)=\alpha(G)$ and $\omega(G-v)=\omega(G)$, for all $v\in V(G)$.
\end{lema}
\begin{prova}
Let $v\in V(G)$. Since $p>1$, $v$ is adjacent to some vertex $u$. Lemmas~\ref{lem:web} and \ref{lema:Santiweb} imply that there is a maximum stable set of $G$ containing $u$. It follows that $\alpha(G-v)=\alpha(G)$. Then, the other equality is a consequence of $\alpha(G)=\omega(\overline G)$. 
\end{prova}

Additionally, the greatest common divisor between $n$ and $\alpha(G)$ plays an important role in our analysis. For arbitrary nonzero integers $a$ and $b$,  the B\'ezout's identity guarantees that the equation $ax+by=\gcd(a,b)$ has an infinity number of integer solutions $(x,y)$. 
As there always exist solutions with positive $x$,
we can define
$$t(a,b)=\min\left\{t\in \N: \frac{at-\gcd(a,b)}{b}\in \Z \right\}.$$
For our purposes, it is sufficient to consider $a$ and $b$ as positive integers.

\begin{lema} \label{lema:t}
Let $a,b \in \N$. If $\gcd(a,b)=b$, then $t(a,b)=1$. Otherwise, $0<t(a,b)<\frac{b}{\gcd(a,b)}$.
\end{lema}
\begin{prova}
If $\gcd(a,b) = b$, then we clearly have $t(a, b) = 1$. Now, assume that $\gcd(a,b)\neq b$. Define the coprime integers $a'=a/\gcd(a,b)$ and $b'=b/\gcd(a,b)>1$. We have that $t(a,b)=t(a',b')$ because $\gcd(a',b')=1$ and $\frac{at - \gcd(a, b)}{b} = \frac{a't - 1}{b'}$, for all $t \in \N$.   
By the B\'ezout's identity, there are integers $x>0$ and $y$ such that $a'x+b'y=1$. Take $t=x \mod b'$, that is, $t=x-\lf \frac{x}{b'} \rf b'$. Therefore, $0\leq t<b'$ and $\frac{ta'-1}{b'}=\left(-y-\lf \frac{x}{b'}\rf a'\right)\in \Z$. Actually, $t>0$ since $b'>1$. These properties of $t$ imply that $0<t(a,b)=t(a',b')\leq t< b'$.
\end{prova}

\subsection{Web}
In this subsection, Theorem~\ref{teo:web} is used to determine the $k$-chromatic number of the graph obtained by removing a vertex from $\web$. For the ease of notation, along this subsection let $t^\star=t(n,p)=t(n,\alpha(\web))$.

\begin{lema} \label{lema:web-v}
For every $k\in \N$ and every vertex $v\in V(\web)$, 
$$
\chi_{k}(\web-v) = 
\left\{
\begin{array}{ll}
\lc \frac{kn}{p} \rc, & \text{if } \gcd(n,p)\neq 1,\\*[0.3cm]
\lc \frac{kn-\lf\frac{k}{t^\star}\rf}{p}\rc, & \text{if } \gcd(n,p)=1,
\end{array}
\right.
$$
\end{lema}
\begin{prova}
Let $q=\gcd(n,p)$. First, suppose that $q>1$. Using Lemma~\ref{lema:trotter}, it is easy to verify that $W^{n/q}_{p/q}$ is a sugbraph of $\web-v$. By Theorem~\ref{teo:web}, we have that 
$$
\chi_k(\web-v)\geq \lc \frac{\frac{n}{q} k}{\frac{p}{q}} \rc = \lc \frac{nk}{p}\rc.
$$  
The converse inequality follows as a consequence of $\chi_k(\web-v)\leq \chi_k(\web)$. 

Now, assume that $q=1$. 

\begin{claim} \label{claim:UB}
$\chi_k(\web-v)\leq \lc \frac{nk - \lf \frac{k}{t^\star} \rf}{p} \rc$.
\end{claim}
\begin{prova}
By the symmetry of $\web$, we only need to prove the statement for $v=v_{n-1}$. 
Since $q=1$, $p$ divides $nt^\star -1$. Let us use (\ref{eq:Sweb}) to define $\Xi = \langle S_{0}, S_{p}, \ldots, S_{\left(\frac{nt^\star-1}{p} - 1\right) p} \rangle$, which is a sequence (modulus $n$) of integer numbers starting at $0$ and ending at $n-2$. Notice that it covers $t^\star$ times each integer from $0$ to $n-2$. Using this sequence $\lf \frac{k}{t^\star}\rf$ times, we get a $\left( \lf \frac{k}{t^\star}\rf t^\star \right)$-fold coloring of $\web-v$ with $\frac{nt^\star -1}{p}\lf \frac{k}{t^\star}\rf$ colors. If $t^\star$ divides $k$, then we are done. Otherwise, by Theorem~\ref{teo:web} and the fact that $\web-v\subseteq \web$, we can have an additional $\left( k-\lf \frac{k}{t^\star}\rf t^\star \right)$-fold coloring with at most $\lc \frac{n}{p}\left(k-\lf \frac{k}{t^\star}\rf t^\star \right)\rc$ colors. Therefore, we obtain a $k$-fold coloring with at most $\frac{nt^\star -1}{p}\lf \frac{k}{t^\star }\rf +\lc \frac{n}{p}\left(k-\lf \frac{k}{t^\star }\rf t^\star \right)\rc= \lc \frac{nk - \lf \frac{k}{t^\star } \rf}{p} \rc$ colors.
\end{prova}

\begin{claim} \label{claim:LB}
$\chi_k(\web-v)\geq \lc \frac{(nt^\star-1)k}{pt^\star}\rc$
\end{claim}
\begin{prova}
By Theorem~\ref{teo:web}, it suffices to show that $W^{n'}_{t^\star}$ is a web included in $\web-v$, where $n'=\frac{nt^\star-1}{p}\in \Z$ because $q=1$. By Lemma~\ref{lema:t}, $t^\star<p$ implying that $n'<n$. Therefore, we only need to show that $W^{n'}_{t^\star}$ is a subgraph of $\web$. First, notice that $n\geq 2p+1$ and so $n'\geq 2t^\star + \frac{t^\star-1}{p}\geq 2t^\star$. Thus, $W^{n'}_{t^\star}$ is indeed a web. To show that it is a subgraph of $\web$, we apply Lemma~\ref{lema:trotter}. On one hand, $nt^\star\geq nt^\star-1=n'p$. On the other hand, $n(t^\star-1)\leq n'(p-1)$ if, and only if, $n'\leq n-1$. Therefore, the two conditions of Lemma~\ref{lema:trotter} hold.
\end{prova}

By claims~\ref{claim:UB} and \ref{claim:LB}, we get
$$
 \lc \frac{nk - \lf \frac{k}{t^\star} \rf}{p} \rc \geq \chi(\web-v)\geq  \lc \frac{nk -  \frac{k}{t^\star}}{p} \rc. 
$$
To conclude the proof, we show that equality holds everywhere above.
Let us write $k=\lf \frac{k}{t^\star}\rf t^\star + r$, where $0\leq r< t^\star$. By the definition of $t^\star$, we have that  $\frac{nt^\star -1}{p}\in \Z$ but $\frac{nr-1}{p}\notin \Z$. It follows that
\begin{multline*}
 \lc \frac{nk -  \frac{k}{t^\star}}{p} \rc \geq   \lc \frac{nk - \lf \frac{k}{t^\star} \rf-1}{p} \rc = \frac{nt^\star-1}{p}\lf \frac{k}{t^\star} \rf + \lc \frac{nr-1}{p} \rc = \\
\lc \frac{nt^\star-1}{p}\lf \frac{k}{t^\star} \rf + \frac{nr}{p} \rc =\lc \frac{nk - \lf \frac{k}{t^\star} \rf}{p} \rc. 
\end{multline*}
 \end{prova}

\begin{observacao}\label{obs:facil}
The proof of Lemma~\ref{lema:web-v} provides the alternative equality $\chi_k(\web-v)=\lc \frac{kn-\frac{k}{t^\star}}{p}\rc$ when $\gcd(n,p)=1$.
\end{observacao}

Removing a vertex from a graph may decrease its $k$-th chromatic number of a value varying from $0$ to $k$. For webs, the expressions of $\chi_{k}(\web)$ and $\chi_{k}(\web-v)$ given above together with Lemma~\ref{lemaTetoPiso} bound this decrease as follows.

\begin{corolario}\label{cor:web-v}
Let $k\in \N$ and $v\in V(\web)$. If $\gcd(n,p)\neq 1$, then $\chi_k(\web)=\chi_{k}(\web-v)$. Otherwise, 
$\lf \frac{k}{pt^\star} \rf \leq \chi_k(\web)-\chi_{k}(\web-v) \leq \lc \frac{k}{pt^\star} \rc$.
\end{corolario}

\begin{observacao}
An important feature of a $\chi$-critical graph $G$ is that, for every vertex $v\in V(G)$, there is always an optimal coloring where $v$ does not share its color with the other vertices. Such a property makes it easier to show that inequalities based on $\chi$-critical graphs are facet-defining for $1$-fold coloring polytopes \cite{CamCorFro04,MenDia08,Pal10}. For $k\geq 2$, Corollary~\ref{cor:web-v} establishes that cliques are the unique webs for which there exists an optimal $k$-fold coloring where a vertex does not share any of its $k$ colors with the other vertices. Indeed, for $p\geq 2$ and $k\geq 2$, the upper bound given in Corollary~\ref{cor:web-v} leads to $\chi_k(\web)-\chi_{k}(\web-v) \leq \lc \frac{k}{2} \rc<k$.
\end{observacao}

Next, we identify the values of $n$, $p$, and $k$ for which the lower bound given in Corollary~\ref{cor:web-v} is nonzero. In other words, we characterize the $\chi_k$-critical webs, for every $k\in \N$.

\begin{teorema}\label{teo:critweb}
Let $k\in \N$. If $\gcd(n,p)\neq 1$, then $\web$ is not $\chi_k$-critical. Otherwise, the following assertions are equivalent:
\begin{enumerate}[(i)]
\item \label{it:web1}
$\web$ is $\chi_k$-critical;
\item \label{it:web2}
$k\geq  p t^\star$ or $0<\frac{nk}{p}-\lf \frac{nk}{p}\rf \leq \frac{k}{p
t^\star}$;
\item \label{it:web3}
$k\geq  pt^\star $ or $k=at^\star + bp$ for some integers $a\geq 1$ and $b\geq
0$.
\end{enumerate}
\end{teorema}
\begin{prova}
The first part is an immediate consequence of Corollary~\ref{cor:web-v}. For the second part, assume
that $\gcd(n,p)=1$, which means that $\frac{nt^\star-1}{p}\in \Z$. Let $r=kn \mod p$, i.e. $ \frac{r}{p} = \frac{kn}{p} - \lf \frac{kn}{p} \rf.$
So, assertion~(\ref{it:web2}) can be rewritten as 
\begin{equation}\label{eq:condweb}
k\geq  p t^\star \text{ or } k\geq  r t^\star  \text{ with } r>0.
\end{equation}
On the other hand, by Theorem~\ref{teo:web} and Remark~\ref{obs:facil}, it follows that
$$
\chi_k(\web) =  \lf \frac{kn}{p}\rf +\lc \frac{r}{p} \rc \quad \text{and} \quad 
\chi_k(\web-v) =  \lf \frac{kn}{p}\rf +\lc \frac{r-\frac{k}{t^\star}}{p} \rc.
$$
Therefore, $\web$ is $\chi_k$-critical if, and only if, $\lc \frac{r}{p} \rc>\lc
\frac{r-\frac{k}{t^\star}}{p} \rc$. If $r=0$, this means that $\lc
\frac{-\frac{k}{t^\star}}{p} \rc\leq -1$ or, equivalently, $k\geq  p t^\star$.
If $r\geq 1$, then the condition is equivalent to $\lc \frac{r-\frac{k}{t^\star}}{p}
\rc \leq 0$ or still $k\geq  r t^\star$. As $r<p$, we can conclude that $\web$ is $\chi_k$-critical if, and only if,
condition \eqref{eq:condweb} holds.

To show that \eqref{eq:condweb} implies assertion (\ref{it:web3}), it suffices to show that $k\geq
rt^\star$ and $r>0$ imply that there exist integers $a\geq 1$ and
$b\geq 0$ such that $k=at^\star + bp$. Indeed, notice that $\frac{kn-r}{p}\in \Z$. Then, $\frac{knt^\star-rt^\star}{p}=\frac{(nt^\star-1)k+(k-rt^\star)}{p}\in \Z$. We can deduce that $\frac{k-rt^\star}{p}\in \Z$ or, equivalenty, $k=rt^\star+bp$ for some $b\in \Z$. Since $k\geq rt^\star$ and $r\geq 1$, the desired result follows.

Conversely, let us assume that $k=at^\star + bp$ for some integers $a\geq 1$ and
$b\geq 0$. If $a \geq p$, then we trivially get condition \eqref{eq:condweb}. So, assume that
$a<p$. We claim that $r=a$. Indeed,
$$r=(nat^\star) \mod p=nat^\star- \lf \frac{(nt^\star-1)a+a}{p}\rf p = a - \lf
\frac{a}{p}\rf p = a.$$ 
Since $a\geq 1$ and $b\geq 0$, we have that $k\geq rt^\star$ and $r>0$. 
\end{prova}

As an immediate consequence of Theorem~\ref{teo:critweb}(\ref{it:web3}), we have the characterization of $\chi_*$-critical webs.
\begin{teorema}\label{teo:webx*}
The following assertions are equivalent:
\begin{enumerate}[(i)]
\item \label{it:web*1}$\web$ is $\chi_*$-critical;
\item \label{it:web*2}$\web$ is $\chi$-critical;
\item \label{it:web*3} 
$\alpha(\web)$ divides $n-1$. 
\end{enumerate}
\end{teorema}
\begin{prova}
Since any $\chi_*$-critical graph is $\chi$-critical, we only need to prove that (\ref{it:web*2}) implies (\ref{it:web*3}), and (\ref{it:web*3}) implies (\ref{it:web*1}). Moreover, (\ref{it:web*3}) is equivalenty to $t^\star=\gcd(n,p)=1$. To show the first implication, we apply Theorem~\ref{teo:critweb}(\ref{it:web3}) with $k=1$. It follows that $\gcd(n,p)=1$ and $at^\star\leq 1$ for $a\geq 1$. Therefore, $t^\star=\gcd(n,p)=1$. For the second part, notice that any $k\in \N$ can be written as $k=at^\star + bp$ for $a=k\geq 1$ and $b=0$, whenever $t^\star=1$. The result follows again by Theorem~\ref{teo:critweb}(\ref{it:web3}). 
\end{prova}

\begin{corolario}
Cliques, odd holes and odd anti-holes are all $\chi_*$-critical.
\end{corolario}

\subsection{Antiwebs}
Now, we turn our attention to $\antiweb$. Similarly to the previous subsection, Theorem~\ref{teo:antiweb} is used to determine the $k$-chromatic number of the graph obtained by removing a vertex from $\web$.  In this subsection, let $t^\star=t(n,\alpha(\antiweb))$.

\begin{lema} \label{lema:antiweb-v}
For every $k\in \N$ and every vertex $v\in V(\antiweb)$, 
$$\chi_{k}(\antiweb-v) = 
\left\{
\begin{array}{ll}
\lc \frac{kn}{\alpha(\antiweb)} \rc & \text{if } \gcd(n,p)=p,\\*[0.3cm]
\lc \frac{k(n-1)}{\alpha(\antiweb)} \rc & \text{if } \gcd(n,p)\neq p.
\end{array}
\right.
$$
\end{lema}
\begin{prova}
First assume that $p$ divides $n$. Using Lemma~\ref{lem:up} and Corollary~\ref{cor:eq}, we get
$$
k\omega(\antiweb-v)\leq \chi_k(\antiweb-v)\leq \chi_k(\antiweb)=k\omega(\antiweb).
$$
By Lemma~\ref{lem:eq}, $\omega(\antiweb)=\omega(\antiweb-v)$ if $p>1$. The same equality trivially holds when $p=1$ since $\overline{W}^n_1$ has no edges. These facts and the above expression show that $\chi_k(\antiweb-v)=\chi_k(\antiweb)=\lc \frac{kn}{\alpha(\antiweb)}\rc$.

Now assume that $\gcd(n,p)\neq p$. Then, $p>1$ and $n>2p$. By lemmas~\ref{lema:lex} and \ref{lem:eq}, we have that $\chi_{k}(\antiweb-v) \geq  \lc \frac{k(n-1)}{\alpha(\antiweb)} \rc$. Now, we claim that $\antiweb-v$ is a subgraph of $\overline{W}^{n-1}_p$. First, notice that this antiweb is well-defined because $n-1\geq 2p$. Now, let $v_iv_j\in E(\antiweb-v)\subset E(\antiweb)$. Then $|i-j|<p$ or $|i-j|>n-p>(n-1)-p$. Therefore, $v_iv_j\in E(\overline{W}^{n-1}_p)$. This proves the claim. Then, Theorem~\ref{teo:web} implies that $\chi_{k}(\antiweb-v) \leq  \chi_{k}(\overline{W}^{n-1}_p)=\lc \frac{k(n-1)}{\alpha(\overline{W}^{n-1}_p)} \rc$. Moreover, since $p$ does not divide $n$, it follows that $\alpha(\overline{W}^{n-1}_p)=\lf \frac{n-1}{p} \rf= \lf \frac{n}{p} \rf=\alpha(\overline{W}^{n}_p)$. This shows the converse inequality $\chi_{k}(\antiweb-v) \leq  \lc \frac{k(n-1)}{\alpha(\antiweb)} \rc$.
\end{prova}

Using again Lemma~\ref{lemaTetoPiso}, we can now bound the difference between $\chi_{k}(\antiweb)$ and $\chi_{k}(\antiweb-v)$.
 
\begin{corolario} \label{cor:antiweb-v}
Let $k\in \N$ and $v\in V(\antiweb)$. If $p$ divides $n$, then $\chi_k(\antiweb-v)=\chi_k(\antiweb)$. Otherwise, $\lf \frac{k}{\alpha(\antiweb)} \rf \leq \chi_k(\antiweb)-\chi_{k}(\antiweb-v) \leq \lc \frac{k}{\alpha(\antiweb)} \rc$.
\end{corolario}

\begin{observacao}
For $k\geq 2$, no antiweb has an optimal $k$-fold coloring where a vertex does not share any of its $k$ colors with other vertices. Since $\alpha(\antiweb)\geq 2$, Corollary~\ref{cor:antiweb-v} establishes that $\chi_k(\antiweb)-\chi_{k}(\antiweb-v) \leq \lc \frac{k}{2}\rc <k$, whenever $k\geq 2$.
\end{observacao}

The above results also allow us to characterize the $\chi_k$-critical antiwebs, as follows.

\begin{teorema}\label{teo:critant}
Let $k\in \N$. If $\gcd(n,p)=p$, then $\antiweb$ is not $\chi_k$-critical. Otherwise, the following assertions are equivalent:
\begin{enumerate}[(i)]
\item \label{it:ant1}
$\antiweb$ is $\chi_k$-critical;
\item \label{it:ant2}
$k\geq \alpha(\antiweb)$ or $0<\frac{nk}{\alpha(\antiweb)}-\lf \frac{nk}{\alpha(\antiweb)}\rf \leq \frac{k}{\alpha(\antiweb)}$;
\item \label{it:ant3}
$k\geq \alpha(\antiweb)$ or $k=at^\star + bq$ for some integers $a\geq 1$ and $b\geq \frac{a(\gcd(n,\alpha(\antiweb))-t^\star)}{q}$, where $q=\alpha(\antiweb)/\gcd(n,\alpha(\antiweb))$.
\end{enumerate}
\end{teorema}
\begin{prova}
We use Theorem~\ref{teo:antiweb} and Lemma~\ref{lema:antiweb-v} to get the expressions of $\chi_k(\antiweb)$ and $\chi_k(\antiweb-v)$. Then, the first part of the statement immediately follows. Now assume that $\gcd(n,p)\neq p$. Let $\alpha=\alpha(\antiweb)$ and $r=kn \mod \alpha$ so that $\frac{r}{\alpha} = \frac{kn}{\alpha} - \lf \frac{kn}{\alpha} \rf.$
It follows that
$$
\chi_k(\antiweb) = \lf \frac{kn}{\alpha}\rf +\lc \frac{r}{\alpha} \rc \quad \text{and} \quad
\chi_k(\antiweb-v) = \lf \frac{kn}{\alpha}\rf +\lc \frac{r-k}{\alpha} \rc.
$$
Therefore, $\web$ is $\chi_k$-critical if, and only if, $\lc \frac{r}{\alpha} \rc>\lc
\frac{r-k}{\alpha} \rc$. If $r=0$, this means that $\lc
-\frac{k}{\alpha}\rc\leq -1$ or, equivalently, $k\geq  \alpha$.
If $r\geq 1$, then the condition is equivalent to $\lc \frac{r-k}{\alpha}
\rc \leq 0$ or still $k\geq  r$. As $r<\alpha$, we can conclude that $\web$ is $\chi_k$-critical if, and only if,
\begin{equation}\label{eq:condantiweb}
k\geq  \alpha, \text{ or } k\geq  r \text{ and } r>0.
\end{equation}
Notice that this is exactly assertion~(\ref{it:ant2}). 

To show the remaining equivalence, we use again \eqref{eq:condantiweb}. Let $g=\gcd(n,\alpha)$. By the definitions of $r$ and $t^\star$, we have that 
$\frac{gk-rt^\star}{\alpha}=\frac{nk-r}{\alpha}t^\star-\frac{nt^\star-g}{\alpha}k\in \Z$. It follows that $k=at^\star+bq$ for some $b\in \Z$ and $a=r/g\in \Z$. Therefore, the second alternative of \eqref{eq:condantiweb} implies the second alternative of assertion (\ref{it:ant3}). This leads to one direction of the desired equivalence.

Conversely, let us assume that assertion (\ref{it:web3}) holds, that is, there exist integers $a\geq 1$ and $b$ such that $k=at^\star + bq$ and $bq\geq ag-at^\star$. Then, $k\geq ag$. If $ag \geq \alpha$, then we trivially get item (\ref{it:web2}). So, assume that $ag< \alpha$. We will show that $r=ag$. Indeed,
\begin{multline*}
r=(nat^\star + \frac{nb}{g}\alpha) \mod \alpha=(nat^\star) \mod \alpha= \\
nat^\star- \lf \frac{(nt^\star-g)a+ag}{\alpha}\rf \alpha = ag - \lf \frac{ag}{\alpha}\rf \alpha = ag.
\end{multline*}
Since $a\geq 1$ and $k\geq ag$, we have that $k\geq
r$ and $r>0$, showing the converse implication.
\end{prova}

The counterpart of Theorem~\ref{teo:webx*} for antiwebs can be stated now.
\begin{teorema}\label{teo:antx*}
The following assertions are equivalent:
\begin{enumerate}[(i)]
\item \label{it:1*} $\antiweb$ is $\chi_*$-critical;
\item \label{it:2*} $\antiweb$ is $\chi$-critical;
\item \label{it:3*} 
$\alpha(\antiweb)$ divides $n-1$. 
\end{enumerate}
\end{teorema}
\begin{prova}
Let $\alpha=\alpha(\antiweb)$, $g=\gcd(n,\alpha)$ and $q=\alpha/g$. It is trivial that (\ref{it:1*}) implies (\ref{it:2*}). Now assume that $\antiweb$ is $\chi$-critical. By applying Theorem~\ref{teo:critant}(\ref{it:ant3}) with $k=1$, we have that $at^\star+bq=1$ and  $bq\geq ag-at^\star$, for some integers $a\geq 1$ and $b$. Then, $ag\leq 1$. It follows that $a=g=1$ and $b=\frac{1-t^\star}{q}\in \Z$. Since $\gcd(n,p)\neq p$, due to Theorem~\ref{teo:critant}, and $1 \leq t^\star < q$, due to Lemma~\ref{lema:t}, we obtain that $0\geq b\geq \lc \frac{1-q}{q}\rc = 0$. Therefore, $t^\star=g=1$ showing that $\alpha$ divides $n-1$. 

Conversely, assume that $\frac{n-1}{\alpha}\in \Z$, i.e. $t^\star=g=1$. Then, $\alpha\neq \frac{n}{p}$, which implies that $\gcd(n,p)\neq p$. Moreover, any $k\in \N$ can be written as $k=at^\star+bq$ for $a=k\geq 1$ and $b=0$. Since $a$ and $b$ satisfy the conditions of Theorem~\ref{teo:critant}(\ref{it:ant3}), $\antiweb$ is $\chi_*$-critical. 
\end{prova}

\begin{corolario}
If $\gcd(n,p)=1$, then $\antiweb$ is $\chi_*$-critical.
\end{corolario}
\begin{prova}
If $\gcd(n,p)=1$, then $\alpha(\antiweb)=\frac{n-1}{p}$. Since $\frac{n-1}{\alpha(\antiweb)}=p\in \Z$, the result follows by Theorem~\ref{teo:antx*}.
\end{prova}

\bibliographystyle{abbrv}
\bibliography{tcs}

\begin{thebibliography}{10}

\bibitem{BarGaf89}
V.~C. Barbosa and E.~Gafni.
\newblock Concurrency in heavily loaded neighborhood-constrained systems.
\newblock {\em ACM Transactions on Programming Languages and Systems},
  11:562--584, 1989.

\bibitem{CamCorFro04}
M.~Camp{\^e}lo, R.~Corr{\^e}a, and Y.~Frota.
\newblock Cliques, holes and the vertex coloring polytope.
\newblock {\em Information Processing Letters}, 89(4):159--164, 2004.

\bibitem{CheVri02A}
E.~Cheng and S.~de~Vries.
\newblock Antiweb-wheel inequalities and their separation problems over the
  stable set polytopes.
\newblock {\em Mathematical Programming}, 92:153--175, 2002.

\bibitem{CheVri02B}
E.~Cheng and S.~de~Vries.
\newblock On the facet-inducing antiweb-wheel inequalities for stable set
  polytopes.
\newblock {\em SIAM Journal on Discrete Mathematics}, 15(4):470--487, 2002.

\bibitem{ChuRobSeyTho06}
M.~Chudnovsky, N.~Robertson, P.~Seymour, and R.~Thomas.
\newblock The strong perfect graph theorem.
\newblock {\em Annals of Mathematics}, 164:51--229, 2006.

\bibitem{EOSV08}
F.~Eisenbrand, G.~Oriolo, G.~Stauffer, and P.~Ventura.
\newblock The stable set polytope of quasi-line graphs.
\newblock {\em Combinatorica}, 28:45--67, 2008.

\bibitem{GalSas97}
A.~Galluccio and A.~Sassano.
\newblock The rank facets of the stable set polytope for claw-free graphs.
\newblock {\em Journal of Combinatorial Theory, Series B}, 69(1):1--38, 1997.

\bibitem{GelSta75}
D.~Geller and S.~Stahl.
\newblock The chromatic number and other functions of the lexicographic
  product.
\newblock {\em Journal of Combinatorial Theory, Series B}, 19:87--95, 1975.

\bibitem{GilTro81}
R.~Giles and L.~E. Trotter.
\newblock On stable set polyhedra for ${K}_{1,3}$-free graphs.
\newblock {\em Journal of Combinatorial Theory, Series B}, 31(3):313--326,
  1981.

\bibitem{HalKor02}
M.~Halld{\'o}rsson and G.~Kortsarz.
\newblock Tools for multicoloring with applications to planar graphs and
  partial $k$-trees.
\newblock {\em Journal of Algorithms}, 42:334--366, 2002.

\bibitem{HanLabSch09}
P.~Hansen, M.~Labbé, and D.~Schindl.
\newblock Set covering and packing formulations of graph coloring: Algorithms
  and first polyhedral results.
\newblock {\em Discrete Optimization}, 6(2):135--147, 2009.

\bibitem{HerHer02}
F.~Herrmann and A.~Hertz.
\newblock Finding the chromatic number by means of critical graphs.
\newblock {\em ACM Journal of Experimental Algorithmics}, 7:1--9, 2002.

\bibitem{KlaMorPer08}
R.~Klasing, N.~Morales, and S.~Pérennes.
\newblock On the complexity of bandwidth allocation in radio networks.
\newblock {\em Theoretical Computer Science}, 406(3):225--239, 2008.

\bibitem{MenDia08}
I.~Méndez-Díaz and P.~Zabala.
\newblock A cutting plane algorithm for graph coloring.
\newblock {\em Discrete Applied Mathematics}, 156(2):159--179, 2008.

\bibitem{Nar02}
L.~Narayanan.
\newblock {\em Channel Assignment and Graph Multi-coloring}.
\newblock Wiley, 2002.

\bibitem{OpsRob81}
R.~J. Opsut and F.~S. Roberts.
\newblock On the fleet maintenance, mobile radio frequency, task assigment and
  traffic phasing problems.
\newblock In G.~Chartrand, Y.~Alavi, D.~Goldsmith, L.~Lesniak-Foster, and
  D.~Lick, editors, {\em The Theory and Applications of Graphs}, pages
  479--492. Wiley, 1981.

\bibitem{OriSta08}
G.~Oriolo and G.~Stauffer.
\newblock Clique-circulants and the stable set polytope of fuzzy circular
  interval graphs.
\newblock {\em Mathematical Programming}, 115:291--317, 2008.

\bibitem{Pal10}
G.~Palubeckis.
\newblock Facet-inducing web and antiweb inequalities for the graph coloring
  polytope.
\newblock {\em Discrete Applied Mathematics}, 158:2075--2080, 2010.

\bibitem{PecWag06}
A.~Pêcher and A.~K. Wagler.
\newblock Almost all webs are not rank-perfect.
\newblock {\em Mathematical Programming}, 105:311--328, 2006.

\bibitem{GuaYue10}
G.~Ren and Y.~Bu.
\newblock $k$-fold coloring of planar graphs.
\newblock {\em Science China Mathematics}, 53(10):2791--2800, 2010.

\bibitem{Rob79}
F.~S. Roberts.
\newblock On the mobile radio frequency assignment problem and the traffic
  light phasing problem.
\newblock {\em Annals of NY Academy of Sciences}, 319:466--483, 1979.

\bibitem{SchUll97}
E.~R. Scheinerman and D.~H. Ullman.
\newblock {\em Fractional Graph Theory: A Rational Approach to the Theory of
  Graphs}.
\newblock Wiley-Interscience, 1997.

\bibitem{Sta76}
S.~Stahl.
\newblock $n$-tuple colorings and associated graphs.
\newblock {\em Journal of Combinatorial Theory, Series B}, 20:185--203, 1976.

\bibitem{Tro75}
L.~E. Trotter.
\newblock A class of facet producing graphs for vertex packing polyhedra.
\newblock {\em Discrete Mathematics}, 12:373--388, 1975.

\bibitem{Wag04}
A.~K. Wagler.
\newblock Antiwebs are rank-perfect.
\newblock {\em 4OR: A Quarterly Journal of Operations Research}, 2:149--152,
  2004.

\end{thebibliography}

\end{document}